\documentclass[prd,aps,nofootinbib]{revtex4}
\usepackage{amssymb,amsmath,epsfig}
\usepackage[bookmarks]{hyperref}
\usepackage[usenames,dvipsnames]{color}

\newcommand{\proof}{\noindent {\bf Proof. }}

\newtheorem{definition}{Definition}

\newtheorem{proposition}{Proposition}

\newtheorem{theorem}{Theorem}

\begin{document}

\title{On causality violation on a Kerr-de Sitter spacetime}

\author{Thomas Zannias}

\affiliation{
Instituto de F\'\i sica y Matem\'aticas,
Universidad Michoacana de San Nicol\'as de Hidalgo,\\
Edificio C-3, Ciudad Universitaria, 58040 Morelia, Michoac\'an, M\'exico.} 
\email{ zannias@ifm.umich.mx}
\large
\begin{abstract}
\large {The causal properties of the family of Kerr-de Sitter spacetimes are analyzed 
and compared to those of the Kerr family.
First, an inextendible  Kerr-de Sitter spacetime 
is obtained by joining together
Carter's blocks, i.e.\ suitable
four dimensional spacetime regions contained within Killing horizons
or within 
a Killing horizon
 and an asymptotic de Sitter region.
Based on this property, and leaving aside topological identifications,
we show that the causal properties of a Kerr-de Sitter spacetime are 
determined by the causal properties of the individual Carter's blocks
viewed as spacetimes in their own right.
We show that 
any Carter's block
is stably causal 
except for the blocks that 
contain the ring singularity. The latter are vicious
sets, i.e.\  any two events within such block
 can be connected by a future (respectively past) directed timelike curve.
This behavior
is identical to  the causal behavior 
of the Boyer-Lindquist blocks that contain the Kerr ring singularity.
These blocks are also vicious as demonstrated long ago by Carter. 
On the other hand, while for the case
of a naked Kerr singularity the entire spacetime is  vicious and thus closed timelike curves
pass through any event including events in the asymptotic region,
for the case of a Kerr-de Sitter spacetime
the cosmological horizons
protect
the asymptotic de Sitter region 
from  a-causal influences. In that
regard,  a positive cosmological constant
appears to improve the causal behavior of the underlying spacetime.}
 \end{abstract}

\date{\today}


\maketitle
\section{Introduction}
 In this paper, we discuss causality violations taking place within the family of Kerr-de Sitter 
spacetimes. Since the specification of the regions
where these violations take place
 requires 
an understanding of the global structure of the underlying spacetime, 
in this paper we also review some 
of the global properties of the Kerr-de Sitter spacetimes. In that respect, 
we start from a Kerr-de Sitter metric written in Boyer-Lindquist coordinates
and show that this, in general geodesically incomplete, spacetime
referred to as a 
 Carter's 
block\footnote{
A Carter's block is  the analogue of a Boyer-Lindquist block
 employed in
the description of the Kerr (or Kerr-Newman) spacetimes. Since in this work, the background is a Kerr-de Sitter,
and in order to avoid confusion,
we use the term Carter's blocks and further ahead we define precisely these blocks.}, 
can be extended through  Killing horizons to generate
an 
inextendible Kerr-de Sitter spacetime. This property
 permits us to
introduce Carter-Penrose like diagrams and present evidence
suggesting that these
diagrams are similar to
the diagrams describing
 the two dimensional
rotation axis of a Kerr-de Sitter spacetime.
Furthermore, 
we conclude that  the causality properties of any four dimensional  Kerr-de Sitter spacetime
are determined by the causality properties of the 
individual Carter's blocks. We prove that   
 any Carter's block is stably causal 
 except for the blocks that contain the ring singularity.
 These latter  blocks are vicious
sets in the sense defined by Carter \cite{Car3}:
Any two events within a block that contains the ring singularity
can be connected 
 by a future (resp.\ past) directed timelike
 curve.
 This highly counterintuitive 
 property was shown by Carter to 
hold for any Boyer- Lindquist block that contains the ring singularity
in a Kerr (or a Kerr-Newman) spacetime
and in this work we show 
that the
same property holds for 
the blocks that contain 
the ring singularity in any 
Kerr-de Sitter spacetime.\\

In order to prove this property, at
 first we show that
for  any Kerr-de Sitter spacetime
the block that contains
the ring singularity
also contains  a non empty
Carter's time machine (CTM),
i.e.\ a spacetime region 
 where the axial Killing vector field becomes temporal.
Using this CTM, we
prove 
that any two events 
$(I,F)$
in this block
can be joined by  
a timelike future (resp.\ past) pointing curve.
For this, we construct  
 three future pointing timelike segments
 with the following characteristics:
 the first one originates in the event $I$
 and terminates in an event on the equatorial plane $\vartheta =\frac {\pi}{2}$
 of the CTM. The second segment is also timelike and future pointing 
and starts
from the event where the first segment ends
and proceeds with the value of the Boyer Lindquist coordinate $t$ decreasing
while it remains  on the 
 $\vartheta =\frac {\pi}{2}$ plane of the CTM. The final segment,
 is again timelike and future pointing 
 and
 starts 
 from the event where 
  the second segment 
 terminates
 and ends at the event $F$.
 In section $V$ we give the explicit representation
 of these segments
 and discuss their properties.\\

The proof of the vicious nature of any Carter's block
that contains the ring singularity illustrates 
 the role of the CTM in destroying any notion of causality.
 It is worth mentioning here 
 that although Carter in \cite{Car3} strongly emphasizes 
 the negative impact that a non empty CTM has upon 
  causality,
nevertheless in the current literature and standard textbooks
it seems that the vicious nature 
of the blocks that contain the ring singularity
is overlooked.
One gets the impression that
causality violation in the Kerr-Newman family takes place only within the tiny spacetime region that finds itself
within the CTM, while the vicious nature of the entire block that contains the ring singularity
is rarely mentioned. This work shows (and emphasizes) that, either for the case of Kerr or Kerr-de Sitter,
causality is violated within the entire block that contains the ring singularity.\\

The structure of the present paper is as follows: in the next section, we introduce the family of  Kerr-de Sitter metrics. The subsequent section contains 
a brief construction of the maximal analytical extension of the rotation axis of a Kerr-de Sitter spacetime,
while section $IV$ 
discusses the global structure of a four dimensional Kerr-de Sitter.
In section $V$,  we prove 
three propositions
describing the causal properties of a Kerr-de Sitter spacetime.
We finish the paper with a discussion of some open problems
while in an Appendix the reader is reminded of a few basic notions of causality theory.\\

\section{Some Properties of the Kerr-de Sitter Spacetimes }
The family of the Kerr-de Sitter spacetimes was discovered
long ago by  Carter  \cite{Car1},\cite{Car2}. 
In a local set of Boyer-Lindquist coordinates $(t,\varphi,r,\vartheta)$,
the Kerr-de Sitter metric $g$ takes the form:   
\begin{equation}
g = -\frac {\Delta (r)}{I^{2}\rho^{2}}[dt-a\sin^{2}{\theta}d\varphi]^{2}+
\frac {\hat {\Delta}(\vartheta)\sin^{2}\vartheta}{I^{2}\rho^{2}} [ a dt - (r^{2}+a^{2})d\varphi]^{2} 
+\frac {\rho^{2}} {\Delta(r)}dr^{2}+ \frac {\rho^{2}} {\hat {\Delta}(\vartheta)}d\vartheta^{2}
\label{Eq:g}
\end{equation}
$$
\rho^2 := r^2 + a^2\cos^2\vartheta,\quad
\Delta(r) :=-\frac {1}{3}\Lambda r^2(r^{2}+a^{2})+r^{2}- 2m r + a^2,\quad {\hat \Delta}(\vartheta) :=1+\frac {1}{3}\Lambda a^2\cos^{2}\vartheta,\quad I:=1+\frac {1}{3}\Lambda a^2,
$$
where above and hereafter $\Lambda>0$ stands for the cosmological constant,
$m$ is the mass parameter and $a$ is a rotation parameter. The $t$-coordinate takes its  values over the real line, 
the angular coordinates $(\vartheta, \varphi)$ vary in the familiar range, while 
$r$  is restricted to suitable open sets of the real line that are specified further below.
The fields
$\xi_{t}=\frac {\partial }{\partial t}$ and  $\xi_{\varphi}=\frac {\partial }{\partial \varphi}$
are commuting Killing  fields
and the zeros 
of $\xi_{\varphi}$ define the rotation axis.\\
From 
(\ref{Eq:g}), it follows that the non vanishing covariant components $g_{\mu\nu}$
of $g$ are:  
\begin{equation}
 g_{tt}=-\frac {{\Delta({r})}-{{\hat \Delta}}({\vartheta})a^{2}sin^{2}\vartheta}{I^{2}\rho^{2}},~~~
 g_{t\varphi}=\frac {\Delta(r)-{\hat \Delta}({\vartheta})(r^{2}+a^{2})}
{I^{2}\rho^{2}}asin^{2}\vartheta,
 \label{Cov1} 
  \end{equation}
\begin{equation}
g_{\varphi \varphi}=
 \frac {{\hat \Delta}({\vartheta})(r^{2}+a^{2})^{2}-\Delta(r)a^{2}sin^{2}\vartheta}{I^{2}\rho^{2}}sin^{2}\vartheta, ~~~~~
 g_{rr}=\frac {\rho^{2}}{\Delta(r)},\quad\quad g_{\vartheta \vartheta}=\frac {\rho^{2}}{\hat \Delta(\vartheta)}
\label{Cov2}
\end{equation}
while the non vanishing  contravariant components $g^{\mu\nu}$ are:
\begin{equation}
g^{tt}=-\frac {I^{2}[{\hat {\Delta}(\vartheta)}(r^{2}+a^{2})^{2}-{\Delta(r)}a^{2}sin^{2}\vartheta]}{\rho^{2}{\hat {\Delta}(\vartheta)}{\Delta(r)}},~~~~~
g^{t\varphi}=\frac {I^{2}a[{\Delta(r)}-{\hat {\Delta}(\vartheta)}(r^{2}+a^{2})]}{
\rho^{2}{\hat {\Delta}(\vartheta)}{\Delta(r)}}
\label{Eq:AF}
\end{equation}
\begin{equation}
g^{\varphi\varphi}=\frac {I^{2}[{\Delta(r)}-{\hat {\Delta}(\vartheta)}a^{2}sin^{2}\vartheta]}{
\rho^{2}sin^{2}\vartheta{\hat{\Delta}(\vartheta)}{\Delta(r)}},~~~~~ g^{rr}=\frac {\Delta{(r)}}{\rho^{2}},
 \quad  g^{\vartheta\vartheta}=\frac {\hat {\Delta} {(\vartheta)}}{\rho^{2}}. 
 \label{Eq:BBF}
\end{equation}

Algebraic manipulations 
of the scalar invariants show that 
the curvature of  (\ref{Eq:g})
becomes unbounded as $\rho \to 0 $, i.e.\ as the ring  ($r= 0$, $\vartheta= \frac {\pi}{2}$)
is approached. Besides this ring-like curvature singularity,  singularities in
the components of $g$ in (\ref{Eq:g}) occur along 
the rotation
axis, i.e.\ at  $\sin\vartheta=0$, and these  singularities can be 
removed by introducing
 local coordinates
or  introducing generalized Kerr-Schild coordinates.
Singularities in the components of (\ref{Eq:g}) also occur at the roots of the quartic equation $\Delta(r)=0$ 
and these are also coordinate singularities. Depending upon the values of $(\Lambda, m, a),$ the quartic equation $\Delta(r)=0$ may admit up to four distinct real roots
exhibiting one of the following arrangements (see for instance the discussion in \cite{LakZan}):\\

a) all four roots are real and distinct, arranged according to: $r_{1}<0<r_{2}<r_{3}<r_{4}$,\\ 
 
b) all roots are real but $r_{2}$  is doubly degenerate, i.e.
$r_{1}<0<r_{2}=r_{3}<r_{4}$, \\ 
  
c) all roots are real but $r_{4}$ is doubly degenerate, i.e. 
$r_{1}<0<r_{2}<r_{3}=r_{4}$,\\  

d) the three positive roots coincide, i.e. 
$r_{1}<0<r_{2}=r_{3}=r_{4}$,\\  

e) the equation $\Delta(r)=0$ admits a pair of complex conjugate roots\footnote{Note that the possibility that
$\Delta(r)=0$ admits two pairs of complex conjugate roots
is not compatible with $\Lambda>0$ and $a^{2}>0$.} and a pair of real roots: $r_{1}<0<r_{2}.$\\  

Clearly, any open interval
 $(r_{i},r_{i+1})$ 
where $r_{i}$, $r_{i+1}$ 
are consecutive roots of $\Delta (r)=0$, combined  
with the metric $g$ in 
(\ref{Eq:g}) defines
a (geodesically incomplete) spacetime
covered by a single Boyer-Lindquist chart $(t,r,\vartheta,\varphi)$.
Any one of these spacetimes is denoted hereafter by 
 $(T_{(r_{i},r_{i+1})},g)$ and is referred 
as a Carter's block\footnote{It should be mentioned that the rotation axis is also 
considered as being part 
of a Carter's block even though points on this axis are not
covered by the Boyer-Lindquist coordinates.}.
The blocks $(T_{(r_{4},\infty)}, g)$ 
and
 $(T_{ (-\infty, r_{1})}, g)$ 
 define  the two asymptotic blocks, while
$(T _{(r_{1},  r_{2})}, g)$ stands for the block that contains  the ring singularity.
As we shall see in the next sections, Carter's blocks
can be glued  together 
along 
null hypersurfaces that are actually Killing horizons
and thus these
blocks determine
the global structure of a Kerr-de Sitter spacetime.\\


We finish this section by introducing a few 
auxiliary fields
that will be helpful later on. The canonical vector fields:
  \begin{equation}
V=(r^{2}+a^{2})\frac {\partial}{\partial t} +a\frac {\partial}{\partial \varphi},\quad W=\frac {\partial}{\partial \varphi}+a\sin^{2}\vartheta \frac {\partial}{\partial t}
\label{Eq:CF}
\end{equation}
are well defined on any Carter's block
and satisfy:
 \begin{equation}
g(V,V)=-\frac {\rho^{2}\Delta(r)}{I^{2}},\quad g(W,W)=\frac {\rho^{2}\sin^{2}\vartheta{\hat \Delta}(\vartheta)}{I^{2}},\quad
g(V,W)=0.
\label{Eq:BF}
\end{equation}
These fields and the coordinate basis fields in 
(\ref{Eq:g}) obey:
$$
 g(V, \frac {\partial}{\partial r})=g(V, \frac {\partial}{\partial \vartheta})=
g(W, \frac {\partial}{\partial r})=g(W, \frac {\partial}{\partial \vartheta})=0
$$

\begin{equation}
g(\frac {\partial}{\partial r},\frac {\partial}{\partial r})=\frac {\rho^{2}}{\Delta(r)},\quad
g(\frac {\partial}{\partial \vartheta},\frac {\partial}{\partial \vartheta})=\frac {\rho^{2}}{{\hat \Delta}(\vartheta)}.
\label{Eq:RR}
\end{equation}
\\
while the gradient fields,
 \begin{equation}
L_{t}=\nabla^{a}t\frac {\partial}{\partial x^{a}},\quad
L_{r}=\nabla^{a}r\frac {\partial}{\partial x^{a}},
\label{Eq:TF}
\end{equation}
satisfy
\begin{equation}
g(L_{t}, L_{t})=g^{tt}=-\frac {I^{2}[{\hat {\Delta}(\vartheta)}(r^{2}+a^{2})^{2}-{\Delta(r)}a^{2}\sin^{2}\vartheta]}{\rho^{2}{\hat {\Delta}(\vartheta)}{\Delta(r)}},\quad\quad
g(L_{r}, L_{r})=\frac {\Delta(r)}{\rho^{2}}.
\label{Eq:MTF}
\end{equation}
Formulas $(\ref{Eq:CF}-\ref{Eq:MTF})$ 
will be used further ahead.

\section{On the Maximal analytical extension of the rotation axis}

The discussion in the previous section suggests that 
the Carter's blocks should be viewed as open submanifolds in a larger
Kerr-de Sitter manifold and the issue we address in the next sections concerns 
the structure of 
this larger Kerr-de Sitter manifold.\\

We recall that Carter, in ref.\cite{Car3}, obtained the maximal analytical extension of the 
Kerr metric guided by the
maximal analytical extension 
of the two dimensional 
 rotation axis of a Kerr spacetime
combined with the behavior of causal geodesics
on a Kerr background\footnote{The maximal analytical extension of the rotation axis of a Kerr 
spacetime was worked out by Carter in 
\cite{Car4}), while the behavior of orbits on a Kerr-Newman
was considered in 
\cite{Car3}).}. Interestingly, below, we show
 that  the maximal analytical extension of the rotation axis of a Kerr-de Sitter 
offers clues regarding 
 the global structure of the four dimensional Kerr-de Sitter spacetime\footnote{We restrict our attention to 
the case where $\Delta(r)=0$ admits four distinct real roots 
arranged according to 
$r_{1}<0<r_{2}<r_{3}<r_{4}$. Once
the structure of these spacetimes is understood, it is relatively easy to understand the structure of spacetimes where
some of the  roots of  $\Delta(r)=0$ coincide.}.
 In order to make this connection clear, 
at first we briefly discuss the maximal analytical extension 
 of the rotation axis of a Kerr-de Sitter spacetime.\\
 
The rotation axis 
of a Kerr-de Sitter spacetime is identified as a two dimensional closed, totally geodesic submanifold consisting of the zeros of the axial Killing field
 (for a definition
 and properties of totally geodesic submanifolds see ref.\cite{Neil}, page $48$). 
Since the Boyer-Lindquist coordinates  in
(\ref{Eq:g}) do
 not cover this submanifold, we introduce new local coordinates $(x,y)$ via 
$x=\sin\vartheta \cos\varphi$ and $y=\cos\vartheta \sin\varphi$ so that $(t,r, x=y=0)$ defines the rotation axis.
Relative to these $(t,r, x,y)$ coordinates the induced metric $g_{in}$ on the axis 
takes the form
(for details see \cite{FT1}):
 \begin{equation}
g_{in}=-f(r)dt^{2}+\frac {dr^{2}}{f(r)},\quad f(r)=\frac {\Delta(r)}{r^{2}+a^{2}},\quad (t,r)\in R \times (r_{i}, r_{i+1})
\label{Eq:DG}
\end{equation}
where 
a factor of $I=1+\frac {1}{3}\Lambda a^2$ has been absorbed in a redefinition of the Killing time.
 Whenever $\Delta(r)=0$ admits four distinct real roots
 $r_{1}<0<r_{2}<r_{3}<r_{4}$, 
 then  (\ref{Eq:DG}) 
 defines five\footnote{We denote by $(M_{-\infty}, g_{-\infty})$,
 $(M_{1}, g_{1})$, $(M_{2}, g_{2})$, $(M_{3}, g_{3})$, $(M_{4}, g_{4})$ these 
 two dimensional spacetimes 
 and in these spacetimes the coordinate  $r$ takes its values 
 respectively in the intervals:  $(-\infty, r_{1}), (r_{1}, r_{2}), (r_{2},r_{3}), (r_{3},r_{4}), (r_{4},\infty).$ } two dimensional  spacetimes representing disconnected components of the rotation axis.
 These five two dimensional spacetimes can be glued\footnote{For this gluing process, at first each of the 
  two dimensional spacetimes 
  defined by 
  (\ref{Eq:DG})  are mapped conformally 
 either into the interior of a diamond configuration
 or to a half diamond configuration (for details of this mapping see for instance \cite{Wal1},\cite{Chr1},\cite{F}).
 The spacetimes in (\ref{Eq:DG})
defined on 
$ (r_{1},r_{2}), (r_{2},r_{3}), (r_{3},r_{4})$
are mapped into the interior of a diamond configuration, while 
 those defined on 
$(-\infty, r_{1})$ and $ (r_{4},\infty)$
are mapped onto a half of a diamond configuration. Each of these five spacetimes
can be time oriented so that 
for any block where  $\Delta(r)>0$,
the timelike Killing field $X=\frac {\partial}{\partial t}$ (or the alternative $\hat X=-\frac {\partial}{\partial t}$) 
can be chosen to provide the future direction, while 
for any  block with
$\Delta(r)<0$ the timelike field 
$\hat X=-\frac {\partial}{\partial r}$ 
(or the alternative $\hat X=\frac {\partial}{\partial r}$)
provides the future direction.}
 together
yielding eventually the maximal analytical extension of the rotation axis.

\begin{figure}[h]
\centering
\begin{minipage}{.5\textwidth}
  \centering
  \includegraphics[width=.8\linewidth]{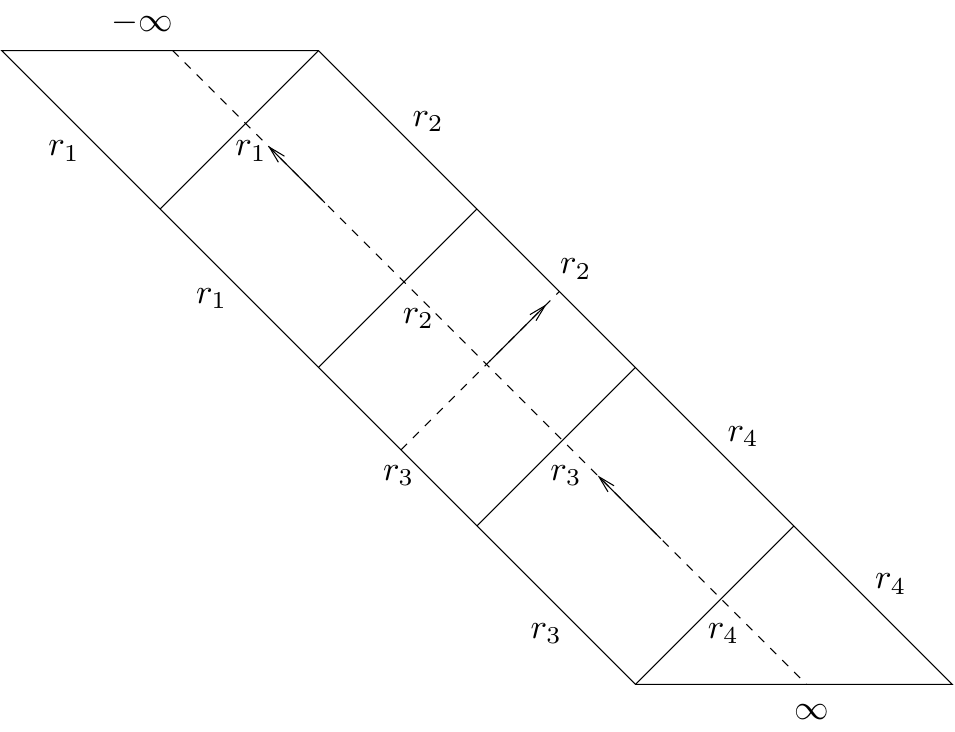}
\end{minipage}%
\begin{minipage}{.5\textwidth}
  \centering
  \includegraphics[width=.8\linewidth]{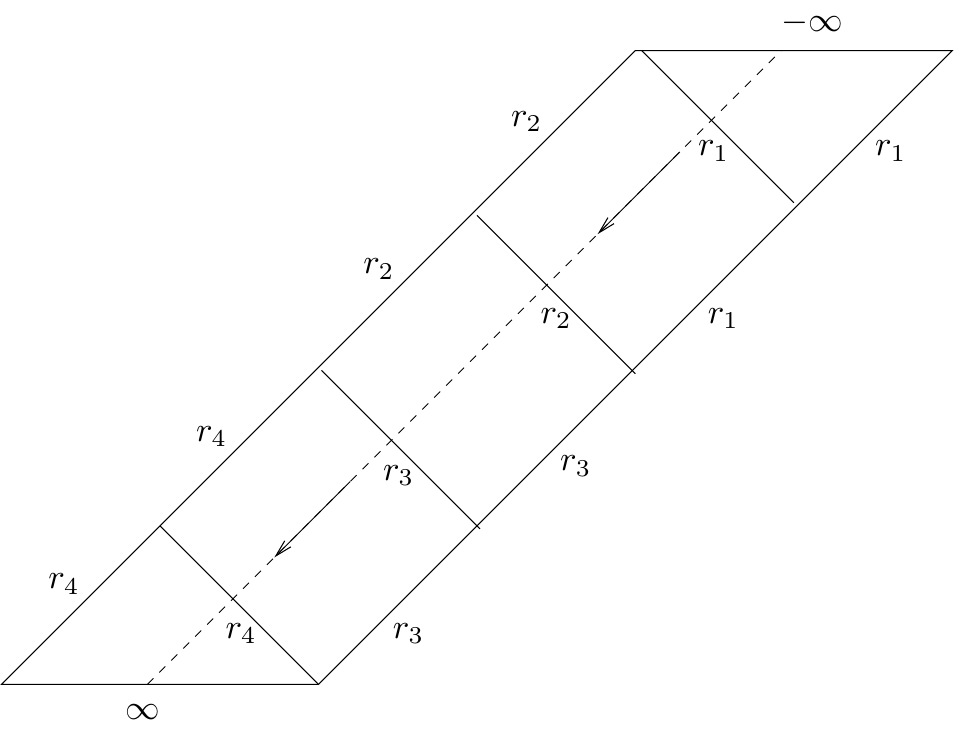}
\end{minipage}
\caption{The figure on the left represents  the Carter-Penrose diagram for the two dimensional ingoing spacetime 
$(IEF, g_{i})$
	 introduced in the main text with the embedding of the five two dimensional 
	 spacetimes included in (\ref{Eq:DG})
	 indicated. Ingoing null geodesics 
	extend
	 from $\infty$
	up to $-\infty$ and thus are complete. By standard conventions these geodesics are considered as
	future pointing and thus time-orient $(IEF, g_{i})$. Incomplete outgoing null geodesics are 
	also indicated.
	The figure on the  right represents the Carter-Penrose diagram
	 for the two dimensional outgoing spacetime $(OEF, g_{o})$. Here, outgoing null geodesics are complete
	and  future pointing running from $-\infty$ to $\infty$. Incomplete ingoing ones
	are also shown.\\ 
As discussed in section IV, these diagrams 
also schematically represent 
the four dimensional 
ingoing Kerr-de Sitter
 $(IKS, \hat g_{i})$
(left figure) and
the four dimensional outgoing Kerr-de Sitter $(OKS, \hat g_{o})$ (right figure). 
In such an interpretation, the blocks 
are four dimensional  Carter's blocks, the null lines marked by $r_{1}, r_{2}$, etc.,
represent Killing horizons. Principal ingoing and outgoing null geodesics are also indicated.
} 
\end{figure}

In order to carry out this gluing processes, we start from $(M_{1},g_{1})$ (see footnote (7) for the definition of $(M_{1},g_{1})$)
and
introduce ingoing Eddington-Finkelstein coordinates  
$(v, \hat r)$ via 
$$dv=dt+\frac {dr}{f(r)},~d\hat r=dr, \quad t\in(-\infty, \infty), ~r\in (r_{1},r_{2}).$$
so that
  \begin{equation}
g_{1}=-f(r)dv^{2}+2dvdr ,\quad (v,r)\in R \times (r_{1},r_{2})
\label{Eq:DGI}
\end{equation}
 where above and whenever there is no danger of confusion we write $r$ instead of $\hat r$.
Since  this $g_{1}$ 
is regular over the roots
of $\Delta(r)=0$, 
using the function $f(r)$ in
(\ref{Eq:DG}), we
 extend  $(M_{1},g_{1})$ 
 by allowing the coordinates $(v,r)$
 to run over $ R \times R$ 
 and refer to this extended spacetime
as a two dimensional ingoing Eddington-Finkelstein, denoted by 
  $(IEF, g_{i})$.  
The extended metric $g_{i}$ is defined by:
$$g_{i}=-f(r)dv^{2}+2dudr ,\quad (v,r)\in R \times R.$$
This
  $(IEF, g_{i})$   
  has the property  that the family of cutves
 $v=const$
 and 
$-r\in (-\infty, \infty)$ 
 represents the ingoing, 
complete family of radial null geodesics  
 with $-r\in (-\infty, \infty)$
 acting as an affine parameter. This null geodesic congruence has 
 $L=- \frac {\partial}{\partial r}$ as the tangent null vector field
and it is customary to consider  $L$ as being future 
pointing
and thus providing the global time orientation on
 $(IEF, g_{i})$. \\

It is not difficult to verify that
any of the five two dimensional  spacetimes included in 
(\ref{Eq:DG}), can be isometrically  embedded
as open submanifolds within  
$(IEF, g_{i})$.
For instance starting from 
$(M_{2},g_{2})$,
 we 
 introduce ingoing Eddington-Finkelstein coordinates $(\hat v,\hat r)$ via 
$$d\hat v=dt+\frac {dr}{|f(r)|},~d\hat r=dr, \quad t\in(-\infty, \infty), ~r\in (r_{2},r_{3})$$
 so that $g_{2}$ takes the form
 \begin{equation}
g_{2}={|f(r)|}dt^{2}-\frac{1}{{|f(r)|}}dr^{2} ={|f(r)|}d{\hat v}^{2}-2d{\hat v}d\hat r ,\quad (\hat v,\hat r)\in R \times (r_{2},r_{3})
\label{Eq:DG2}
\end{equation}
and subsequently embed this $(M_{2},g_{2})$ within 
$(IEF, g_{i})$
 via the map:
  \begin{equation}
\Phi: M_{2}\to IEF : (\hat v, \hat r)\to \Phi(\hat v,\hat r)=(v(\hat v,\hat r), r(\hat v, \hat r))=(-\hat v, \hat r),
\label{Eq:MAP}
\end{equation}
which is  a smooth isometry of $M_{2}$ onto $\Phi(M_{2}).$
For  the case of
 $(M_{3},g_{3})$  the  isometry $\Phi$ has the same form 
   as the one described   
  in  (\ref{Eq:MAP}) with the only exception that $-\hat v$ is replaced by $\hat v$ and so on.
  In view of these embeddings, the conformal Carter-Penrose diagram
 for $(IEF, g_{i})$ has the form shown in the left diagram of Fig.1.\\
  
We now shift our attention
to the outgoing family of null geodesics
and begin considering again  $(M_{1},g_{1})$, but  now introduce outgoing Eddington-Finkelstein coordinates  
$(u, \hat r)$ via: 
$$du=dt-\frac {dr}{f(r)},~d\hat r=dr, \quad t\in(-\infty, \infty), ~r\in (r_{1},r_{2})$$
so that $g_{1}$ takes the form:
 \begin{equation}
g_{1}=-f(r)du^{2}-2dudr ,\quad (u,r)\in R \times (r_{1}, r_{2}).
\label{Eq:DGO}
\end{equation}
Through the same arguments 
 that lead us to
$(IEF, g_{i})$, we now introduce
the outgoing Eddington-Finkelstein spacetime 
$(OEF, g_{o})$   
with
 \begin{equation}
g_{o}=-f(r)du^{2}-2dudr ,\quad (u,r)\in R \times R.
\label{Eq:OGO}
\end{equation}
Clearly 
this $g_{o}$  is regular over the entire domain of the radial coordinate $r$
and for this
$(OEF, g_{o})$, 
the outgoing family of null geodesics is described by $u=const$, $r\in (-\infty, \infty)$
with $r$ acting as an affine parameter. This family
has $L= \frac {\partial}{\partial r}$ as the tangent null vector field
taken to be future 
pointing and thus defines the global time orientation 
on  
$(OEF, g_{o})$.
The remaining two dimensional spacetimes included in
  (\ref{Eq:DG})
can be isometrically embedded as open submanifolds within 
 $(OEF, g_{o})$
 so that the resulting Carter-Penrose diagram is the right diagram shown
  in Fig.1.\\

The final step leading to an extension of
the rotation axis of a Kerr-de Sitter
consists of gluing together the two diagrams shown in Fig.1 in such a manner that
the radial ingoing and outgoing null geodesics become simultaneously complete.
Here, some care is required so that 
the gluing procedure yields an 
extended spacetime 
admitting a consistent time orientation. One way to achieve this
is to start from a copy of an ingoing Eddington-Finkelstein spacetime
$(IEF, g_{i})$
 shown in Fig.1,
and on a specific block introduce simultaneously outgoing Eddington-Finkelstein coordinates.
Subsequently extend that block in the future direction by appending a part of the outgoing
Eddington-Finkelstein spacetime and making sure that the resulting
spacetime admits a consistent time orientation.
Leaving details aside, the resulting Carter-Penrose diagram
is shown in Fig.2 and this diagram is also introduced in refs.\cite{GibHaw1},\cite{Car2}.\\

To finish this section, we  mention that
  the use of 
Eddington-Finkelstein
coordinates 
as the means to construct   
the Carter-Penrose diagram shown in Fig.2,
does not cover
the vertex where the four horizons meet. 
However, this deficiency can be removed by introducing Kruskal coordinates
which are well defined provided the roots of $f(r)=0$ are all simple roots.
We do not enter into these details here (they are discussed in \cite{F},\cite{FT2}),
 but we only mention that the extension shown in Fig.2 
 is a maximal analytical extension of the rotation axis. Maximality follows by 
 verifying that
 any causal geodesic on this two dimensional spacetime is actually complete
 while the analytical nature of the extension follows from the analyticity  of the function $f(r)$
 in (\ref{Eq:DG}).

\begin{figure}
	\includegraphics[width=0.6\textwidth]{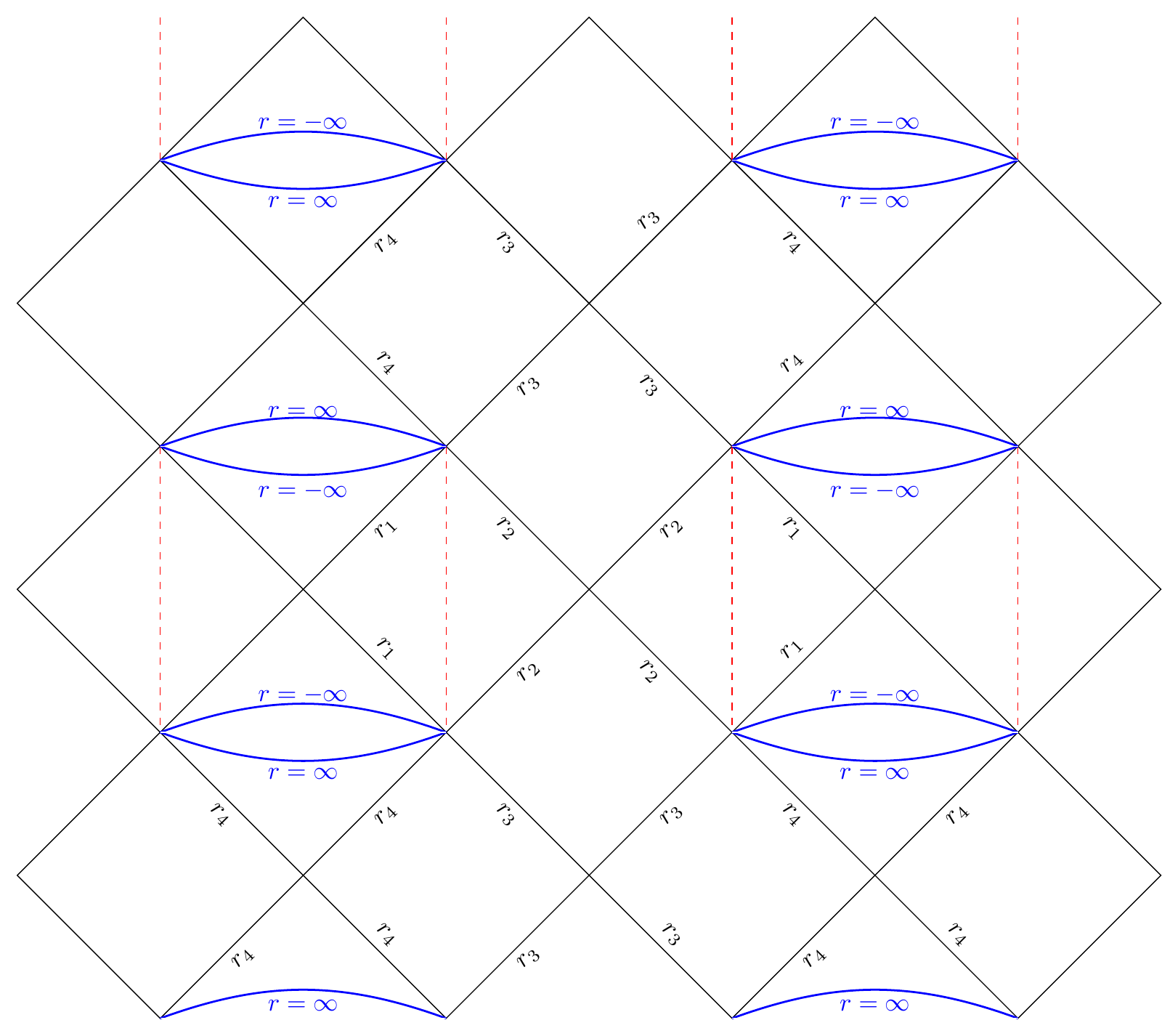}
	\caption{The Carter-Penrose diagram for the two dimensional  rotation axis 
	of a Kerr-de Sitter spacetime (see also \cite{GibHaw1},\cite{Car2}).  
	As we discussed in section IV, this diagram can represent 
	schematically the 
	structure of a four dimensional Kerr-de Sitter spacetime. When interpreted in that way, the dashed          
	lines (red in the colored version of the figure) remind the reader of the location of the ring singularity, 
	while the curved lines (blue in the colored version) representing $r\to \pm  {\infty}$
	signify that these regions are distinct. For this spacetime,
	the causality violating regions are the union of the blocks containing the ring singularity. If the topology 
	of the spacetime is altered by identifying, for instance,  asymptotic regions, then the causality violating regions 
	are altered as well. }

\end{figure}

\section{On the maximal analytical extension of a  Kerr-de Sitter Spacetime}

Even though the construction of the 
maximal  analytical extension of the two dimensional rotation axis of a Kerr-de Sitter spacetime
was a relatively easy task to accomplish, 
the construction of 
the maximal analytical extension of a four dimensional Carter's block  is
not  that straightforward a task\footnote{
As far as we are aware, 
the maximal analytical extension of a four dimension Kerr-de Sitter has not been addressed 
in the literature before. Often and by analogy to what occurs for the Kerr case, the Carter-Penrose diagram 
representing  the axis of a Kerr-de Sitter is interpreted as representing the maximal analytical 
extension of the four dimension Kerr de Sitter. Although it is likely
that is the case, we are not aware of any detailed work supporting
this interpretation. A referee kindly pointed out that some results
that are reported in ref.\cite{FX},
regarding the structure
of the $t=const.$, $r=0$ equatorial disk of a Kerr-de Sitter offer the opportunity for a distinct extension
of the block that contains the ring singularity. Needless to say, issues regarding possible extendability (or extendabilities)  
of Carter's blocks deserve further attention.}.
An extension of a Carter's block could be obtained 
by following
the same method as the one employed 
 by Carter in extending the Kerr
family of metrics (for details see \cite{Car3}). 
However, we should be aware that 
in this  approach, in order to address the maximality property of the extended spacetime,
the behavior of causal geodesics on the extended background
  is required. Although causal
 geodesics on a Kerr-de Sitter
 have been the subject of many investigations, 
these targeted particular families of geodesics, such as 
the family of equatorial \cite{Stuc3}, 
polar \cite{Tso}, spherical \cite{The} or  geodesics confined on a particular 
Carter's block 
\cite{Laz}. In a recent work \cite{FT1}, the  completeness property of geodesics defined  on an
arbitrary 
Carter's block  has been addressed and evidence
was found 
to support
 the view that 
``almost all causal geodesics`` defined initially on a 
Carter's block
can be extended as geodesics through Killing horizons
so they become complete except for those ones
that hit the ring singularity\footnote{Even though we believe
that the results of ref.\cite{FT1} hold for all causal geodesics,
 unfortunately the completeness property of a few families of geodesics needs to be addressed. 
For instance the completeness property
of geodesics hitting the bifurcation spheres, or the completeness
property of 
geodesics through the axis 
have to be worked out. These issues are under investigation and will be reported elsewhere \cite{FT2}.}.
Due to this incomplete understanding of
the behavior of causal geodesics on a Kerr-de Sitter background, 
we outline below an extension of a 
Carter's block  
employing 
a formalism
developed\footnote{The advantage of the extension through the gluing
process advocated in ref.\cite{Neil} lies in the fact that the method does not require
an a priori understanding of the behavior of the causal  geodesics.
Once an extension is obtained, there follows the laborious task
of checking whether 
all causal geodesics 
in the extended spacetime 
  are indeed complete.} in ref. \cite{Neil}.
This formalism is based on the
property 
that
two smooth manifolds  $M$ and $N$ 
admitting two isometric open subsets $(U,V)$ with $U\subset M$ and $V\subset N$,
can be glued 
along these  subsets so that 
a new smooth manifold $Q$ is obtained.
If $\mu: U\to V$ stands for the isometry
between $(U,V)$, then
the resulting manifold $Q$  is defined as the quotient  space  
 $Q=(M \cup N)\setminus \sim$
under a suitable equivalence relation
 $\sim$ spelled out  in 
 \cite{Neil}.
 The proofs of the smoothness, Hausdorff and other properties of the resulting manifold $Q$ 
 are discussed in
ref.\cite{Neil}.\\
 
 To see how this abstract setting applies to 
 the extendability problem of a 
 Carter's block,
 we begin with 
an arbitrary 
$(T_{(r_{i},r_{i+1})}, g)$
and introduce ingoing Eddington-Finkelstein coordinates
 $(v, \overleftarrow{\varphi},r,\vartheta)$  via\footnote{These coordinates
 are based on the family of principal null
 congruences admitted by a Kerr-de Sitter metric. For an introduction to these congruences
 and their role in constructing the Eddington-Finkelstein charts see for instance section $V$ of ref. \cite {FT1}.}:
 \begin{equation}
dv=dt +\frac {I(r^{2}+a^{2})}{{\Delta(r)}}dr,\quad d\overleftarrow{\varphi}=d\varphi+\frac {Ia}{{\Delta(r)}}dr,
\label{Eq:INC}
\end{equation}
so that $g$ in (\ref{Eq:g}) takes the form:
\begin{equation}
g=-\frac {\Delta(r)-a^{2} {\hat {\Delta}({\vartheta})\sin^{2}\vartheta}}{I^{2}\rho^{2}}dv^{2}+\frac {2}{I}dvdr-2 \frac {a}{I}\sin^{2}\vartheta
d\overleftarrow{\varphi}dr-2\frac {a\sin^{2}\vartheta[(r^{2}+a^{2}){\hat \Delta({\vartheta})-{\Delta(r)}}]}
{I^{2}\rho^{2}}dvd\overleftarrow{\varphi}+\nonumber
\end{equation}
\begin{equation}
+\frac {{\rho^{2}}} {{\hat {\Delta}({\vartheta})}}d\vartheta^{2}+\frac {\hat {\Delta}({\vartheta})(r^{2}+a^{2})^{2}-\Delta(r)a^{2}\sin^{2}\vartheta}{I^{2}\rho^{2}}
\sin^{2}\vartheta d\overleftarrow{\varphi}^{2}.
\label{Eq:gINC}
\end{equation}
This $g$ is regular  
across the zeros  of $\Delta(r)=0$
and thus  by letting
the coordinates $(v,r)$
run over the entire real line,
we obtain            
a four dimensional 
spacetime
 $(IKS, \hat g_{i})$
 where 
the extended metric $\hat g_{i}$ is just $g$ in 
 (\ref{Eq:gINC})
 defined now over the extended domain 
 of the $(v,r)$ coordinates.
We refer to this 
  $(IKS, \hat g_{i})$ 
as the 
ingoing Kerr-de Sitter\footnote{Just to stress the formal analogy between
 the extension of the rotation axis and the full four dimensional
 Kerr-de Sitter, this 
 $(IKS, \hat g_{i})$ is the analogue of
 the two dimensional spacetime $(IEF,g_{i})$ introduced in the treatment of the rotation axis.}
   and clearly $(T_{(r_{i},r_{'+1})}, g)$ is an open submanifold of this larger manifold.
Moreover, the map:
\begin{equation}
J: T_{(r_{i}, r_{i+1})}\to J(T_{(r_{i}, r_{i+1})}) \subset IKS: (t, r, \vartheta, \varphi) \to J(t, r,\vartheta,\varphi)=
(r, \vartheta, v(t,r), \overleftarrow{\varphi}(\varphi, r))
\label{Eq:IS1}
\end{equation}
with $(v(t,r), \overleftarrow{\varphi}(\varphi, r))$  the coordinates $(v, \overleftarrow{\varphi})$ 
that 
(\ref{Eq:INC}) assigns to the pair $(t,r)$ and $(\varphi, r)$,
 isometrically
embeds\footnote{The map  $J$, 
plays the role of the
map $\Phi$ defined 
in eq. (\ref{Eq:MAP}),
although here the presence of the coordinate singularity on the  axis 
needs to be given special consideration.
Nevertheless, it can be shown that 
this $J$
has a unique analytical extension as an isometry of 
the entire, i.e. including the axis, $T_{(r_{i}, r_{i+1})}$ into $J(T_{(r_{i}, r_{i+1})})$.}  the 
remaining Carter's block
within  $(IKS, \hat g_{i})$.
This embedding has the  property  that 
the $r=r_{i}$ interfaces become Killing horizons
and 
schematically these embeddings are 
shown in the left diagram in Fig.1, where 
now each  block in that figure should be viewed as a four dimensional region.\\

The ingoing Kerr-de Sitter spacetime $(IKS, \hat g_{i})$ 
has the property that all ingoing principal null geodesics are complete
but the corresponding outgoing ones fail to be so.
In order to achieve completeness of the latter congruence, 
 a different extension of the 
 Carter's block is required. To construct this extension,
 we begin again with 
an arbitrary  block 
$(T_{(r_{i},r_{'+1})}, g)$
but now  introduce outgoing Eddington-Finkelstein coordinates via:
\begin{equation}
du=dt -\frac {I(r^{2}+a^{2})}{{\Delta(r)}}dr,\quad d\overrightarrow{\varphi}=d\varphi-\frac {Ia}{{\Delta(r)}}dr.
\label{Eq:ONC}
\end{equation}
Relative to these 
coordinates, $g$ in (\ref{Eq:g}) takes a form identical to that in (\ref{Eq:gINC}), except 
that 
$(v, \overleftarrow{\varphi})$ are replaced by 
$(u, \overrightarrow{\varphi})$
and the signs in the cross terms $(drdu)$  and 
$(d \overrightarrow{\varphi}dr)$ are now reversed. 
Letting
 $(u, \overrightarrow{\varphi})$ run over the entire real line
 we obtain the four dimensional outgoing Kerr-de Sitter spacetime denoted by 
 $(OKS, \hat g_{o})$.
 Following  the same reasoning
 as 
 for the 
 case of the ingoing Kerr-de Sitter,  the map:
\begin{equation}
\hat J: T_{(r_{i}, r_{i+1})}\to \hat {J}(T_{(r_{i}, r_{i+1})}) \subset OKS: (t, r, \vartheta, \varphi) \to \hat J
(t, r,\vartheta,\varphi)=
(r, \vartheta, u(t, r), \overrightarrow{\varphi}(t,\varphi))
\label{Eq:IS1A}
\end{equation}
with $(u(t,r), \overrightarrow{\varphi}(t, \varphi))$ defined by
(\ref{Eq:ONC}),  isometrically embeds the remaining Carter's block within 
 $(OKS, \hat g_{i})$.
These embeddings are shown 
schematically in the right diagram of Fig.1 where
again the blocks should be viewed as 
four dimensional regions
enclosed between Killing horizons. \\

 The task is now  to 
 assemble the four dimensional 
geodesically  incomplete spacetimes
  $(IKS, \hat g_{i})$
 and 
$(OKS, \hat g_{o})$
in such a manner
 that the resulting extended spacetime
 has the property that both sets of 
principal null congruences 
are complete.
This is not a trivial operation
and this step 
involves 
 the 
gluing process 
 discussed in the section 
$1.4$ of O'Neill's book \cite{Neil}.
To see what is involved, let
$(IKS, \hat g_{i})$
 stand for the manifold\footnote{To make 
matters simple we use the same notation as  the one employed in section
$1.4$ of ref.  \cite{Neil}.}
$M$ and 
 $(OKS, \hat g_{o})$
 for the manifold
 $N$
and, moreover, let $B$ stand for any of the 
$(T_{(r_{i},r_{i+1})}, g)$.
The open submanifolds $J(T_{(r_{i},r_{i+1})})$ of 
$(IKS, \hat g_{i})$
and 
${\hat J}(T_{(r_{i},r_{i+1})})$
of
 $(OKS, \hat g_{o})$
are  isometric via
$$
\mu:={\hat J} (J)^{-1}:
J(T_{(r_{i},r_{i+1})})
\to {\hat J}(T_{(r_{i},r_{i+1})}): (r,\vartheta, v, \overleftarrow{\varphi})\to \mu(r,\vartheta, v, \overleftarrow{\varphi})=
$$
 \begin{equation}
=(r, \vartheta,~ v-2\int^{r}\frac {I(\hat r^{2}+a^{2})}{\Delta({\hat r})}d\hat r, ~\overleftarrow{\varphi}-2\int^{r}\frac {Ia}{\Delta({\hat r})}d\hat r),
\label{Eq:IB}
\end{equation}
and  this isometry $\mu$ provides the important ingredient 
for the gluing process. Via this $\mu$, the spacetimes
 $(IKS, \hat g_{i})$ and 
 $(OKS, \hat g_{o})$ 
 are first glued along the copies 
$U=J(T_{(r_{i},r_{i+1})})$ and $V={\hat J}(T_{(r_{i},r_{i+1})})$ and via this identification
an extension is 
eventually built 
along the same lines as 
the extension of the ``slow`` Kerr  constructed in 
ref.\cite{Neil}.
Although we leave many details to be discussed elsewhere,
we only mention
that the resulting spacetime 
has the property that 
both families of
outgoing and ingoing principal null geodesics
are now complete
and a schematic representation of the global structure is depicted in Fig.2. (see the 
comments in the last paragraph 
of the caption accompanying Fig.1 and also comments in the caption in Fig.2).\\

\section{On the causal properties of Kerr-de Sitter spacetimes }
The discussion of the previous section combined with the diagram\footnote{The maximality property of the
diagram in Fig.2, for the case
that the blocks are considered to be four dimensional,
ought to be worked out in detail and establishing this property is not a trivial task.
 For the rest of this section we will assume that the extension shown in Fig.2 is maximal
and we discuss the consequences of this assumption.} of Fig.2, offers  a view of 
the structure of the family of  Kerr-de Sitter spacetimes
characterized by parameters $(m,a, \Lambda)$ such that  $\Delta(r)=0$ admits four distinct real roots.
In this section, we analyze the causality properties of this
family  and firstly we 
identify the location of the Killing horizons.
 Starting from the ingoing coordinates 
 $(v, \overleftarrow{\varphi}, r,\vartheta)$, 
 the normal vector $N$ of any $r=const$
 hypersurface, has the form: 
\begin{equation}
N={\hat g_{i}}^{\mu\nu}\delta_{\nu}^{r}\frac {\partial}{\partial x^{\mu}}={\hat g_{i}}^{\mu r}\frac {\partial}{\partial x^{\mu}},\quad
x^{\mu}=(v, \overleftarrow{\varphi}, r, \vartheta)
\label{Eq:NKH}
\end{equation}
where ${\hat g_{i}}^{\mu\nu}$  stand for the contravariant components of $g$ 
relative to the ingoing coordinates shown
in (\ref{Eq:gINC}).
Since 
 \begin{equation}
g(N,N)={\hat g_{i}}^{rr}=g^{rr}=\frac {\Delta(r)}{\rho^{2}}
\label{Eq:MNKH}
\end{equation}
it follows that the  set  $r=r_{i}$ defines a null hypersurface\footnote{
The term $\frac {\Delta(r)}{\rho^{2}}$ is well defined over the entire 
domain of validity of the ingoing chart
and this coupled with the fact that the left hand side of 
(\ref{Eq:MNKH}) is an analytic function relative to ingoing 
coordinates shows 
that the claim is not based on Boyer-Lindquist coordinates.
The latter have been used only as an intermediate step.}.
For each 
real root 
$r_{i}$ of $\Delta(r)=0$, we define the constants
\begin{equation}
\Omega_{i}=-\frac {g(\xi_{t},\xi_{\varphi})}{g(\xi_{\varphi},\xi_{\varphi})}=\frac {a}{r_{i}^{2}+a^{2}}.
\label{Eq:ROT}
\end{equation}
and introduce  the Killing fields
\begin{equation}
\hat \xi_{i}=\xi_{t}+\Omega_{i}\xi_{\varphi}=\frac {\partial}{\partial v}+\Omega_{i}\frac {\partial}
{\partial \overleftarrow{\varphi}}
\label{Eq:GKH}
\end{equation}
which 
 become null precisely over the $r=r_{i}$ hypersurfaces.
A computation shows 
that 
\begin{equation}
\nabla^{\mu}[g(\hat \xi_{i},\hat \xi_{i})]=-2k_{i}\hat \xi_{i}^{\mu}
\label{Eq:SGKH}
\end{equation}
which establishes the Killing property of the 
 $r=r_{i}$ hypersurfaces.
 The coefficients 
 $k_{i}$ stand  for the surface gravity\footnote{Our convention for the surface gravity follows the same conventions
 as those in  Wald's  book
ref.\cite{Wald}.}
of the $r_{i}$ horizon
and they are given by (see ref. \cite{FT1}) 
 \begin{equation}
 k_{i}= \frac {1}{2I}\frac{1}{r^{2}_{i}+a^{2}}\frac {\partial \Delta(r)}{\partial r}\Big\vert _{r_{i}}. 
  \label{Eq:SG}
\end{equation}
In the limit of $\Lambda\to 0$  these  $k_{i}$ reduce to the surface gravity for the Killing horizons of 
the Kerr black hole (compare (\ref{Eq:SG})
with the corresponding formulas  for a Kerr black hole in ref.\cite{Wald})
and
moreover, (\ref{Eq:SG}) shows that any Killing horizon corresponding to a double
or higher multiplicity root of $\Delta(r)=0$ is degenerate.
Identical computations based on the outgoing 
 $(u, \overrightarrow{\varphi}, r,\vartheta)$ coordinates shows that the sets $r=r_{i}$ are null hypersurfaces\footnote{
 The reader is warned that the $r=r_{i}$ hypersurfaces defined relative to the 
 the outgoing $(u, \overrightarrow{\varphi}, r,\vartheta)$ coordinates are distinct hypersurfaces
 from those defined by the ingoing 
 $(v, \overleftarrow{\varphi}, r,\vartheta)$ coordinate system.
 For simplicity, we have avoided 
introducing different symbols for the ``radial like'' coordinate in the two systems.}
  and in fact are Killing horizons whose surface gravities  $k_{i}$  are
 still described by (\ref{Eq:SG}).\\
  
 The Killing horizons defined above 
play an important role in determining the causality violating region in any  Kerr-de Sitter
spacetime. Since a Killing horizon is an
achronal set \cite{Neil} (for properties of these sets see   \cite{HE}, \cite{Wald}, \cite{Neil})
 no timelike future directed curve 
 meets a Killing  horizon more than once.
 This property implies that the causal properties of the extended
  Kerr-de Sitter are determined by the causal properties of the 
 Carter's blocks. However the causality properties of these blocks 
can be easily worked out and we begin by 
 first proving the following proposition:
\begin{proposition}
\label{Prop:Mf}
Any Carter's block characterized by $\Delta(r)<0$, is stably causal (see the Appendix for a
brief discussion of stable causality).
\end{proposition}

\proof 
The formulae in (\ref{Eq:RR})
combined with the property $\Delta(r)<0$,
imply that the vector field
$X=\frac {\partial}{\partial r}$ 
is timelike and nowhere vanishing
within the block under consideration
and thus it can time-orient the block.
Moreover,
the gradient 
$L_{r}=\nabla^{a}r\frac {\partial}{\partial x^{a}}$
satisfies
$g(L_{r}, L_{r})=\frac {\rho^{2}}{\Delta(r)}<0$ and thus is timelike. Accordingly, if
$X=\frac {\partial}{\partial r}$ is chosen to
identify the future part of the light cone then 
$\tau=-r$
serves as a time function,
while for the alternative choice, 
i.e.\ if 
$X=-\frac {\partial}{\partial r}$
identifies the future part of the light cone 
then $\tau=r$
serves as a time function.
For any choice,
 all conditions of the Theorem $I$ cited in the Appendix are met and thus any block subject to $\Delta(r)<0$ 
is stably causal.

\begin{proposition}
\label{Prop:MfA}
Any Carter's block with $\Delta(r)>0$
 is stably causal, except for the block that contains the ring singularity.
\end{proposition}

\proof 
From the formulae in (\ref{Eq:BF}),
we have
$g(V,V)=-\frac {\rho^{2}\Delta(r)}{I^{2}}$,
and thus  the vector field $V$
time-orients the block under consideration
(remember the block under consideration does not contain the ring singularity).
In order to construct a time function,
we appeal to the 
gradient field $L_{t}=\nabla^{a}t\frac {\partial}{\partial x^{a}}$
which satisfies:
$$g(L_{t}, L_{t})=g^{tt}=-\frac {I^{2}[{\hat {\Delta}(\vartheta)}(r^{2}+a^{2})^{2}-{\Delta(r)}a^{2}\sin^{2}\vartheta]}{\rho^{2}{\hat {\Delta}(\vartheta)}{\Delta(r)}}.$$
Moreover a computation of the numerator shows:
$$
[{\hat {\Delta}(\vartheta)}(r^{2}+a^{2})^{2}-{\Delta(r)}a^{2}\sin^{2}\vartheta]=
$$
\begin{equation}
=(r^{2}+a^{2})(r^{2}+a^{2}\cos^{2}\vartheta)
+2mra^{2}\sin^{2}\vartheta+
\frac {\Lambda a^{2}}{3}(r^{2}+a^{2})[(r^{2}+a^{2})\cos^{2}\vartheta+r^{2}\sin^{2}\vartheta]
\label{Eq:IR}
\end{equation}
and thus as long as $r>0$,  the right hand side  is positive definite,
which means that 
$L_{t}=\nabla^{a}t\frac {\partial}{\partial x^{a}}$ is everywhere timelike
on  any block 
where $\Delta(r)>0$ and $r>0$.
In addition, from the formulas 
(\ref{Eq:AF}, \ref{Eq:BBF}) and  
 (\ref{Eq:BF}, \ref{Eq:RR})
we find the identity:
$$
L_{t}=-\frac {I^{2}(r^{2}+a^{2})}{\rho^{2}\Delta(r)}V+\frac {I^{2}a}{\rho^{2}{\hat {\Delta}}(\vartheta)}W.
$$
Since $W$ is spacelike,  this identity
shows that $\tau=t$ serves as a time function whenever $X=V$ specifies the
future part of the light cone, while when
$X=-V$ defines the future part, then $\tau=-t$ serves as a time function.
In any case, the proof of the proposition is established by appealing to the theorem $I$ of  the Appendix.\\

We now consider the block that contains the ring singularity.
Even though on this block $\Delta(r)>0$, since now $r$ can take negative values,
 the right hand side of
(\ref{Eq:IR}) fails to be positive definite and thus the argument leading 
to the proof of the proposition $2$ fails.
Instead we have the following proposition:

\begin{proposition}
\label{Prop:MfB}
The  block that contains the ring singularity is totally vicious in the sense of Carter:
Any two events $I,F$ within this  block, can be connected by a future (resp. past) directed timelike
curve lying entirely within the block.
\end{proposition}

\proof The proof of this proposition is long. Firstly, we show that there is a non empty region in this block
where the axial Killing field $\xi_{\varphi}$ becomes timelike,  i.e $g(\xi_{\varphi},\xi_{\varphi})<0$.
This region defines the Carter's time machine\footnote{In the present context,
a time machine is a spacetime region that can generate closed timelike curves
passing through any point in the spacetime under consideration. Here
the region defined in (\ref{Eq:CTM}) acts as a time machine for the block that contains the ring singularity.}
and is denoted hereafter by CTM. Relative to a set of Boyer-Lindquist
coordinates, it is identified as
the set: 
 \begin{equation}
CTM=\left\{ (t,r,\vartheta, \varphi),\quad g(\xi_{\varphi},\xi_{\varphi})=g_{\varphi\varphi}< 0  \right\}.
\label{Eq:CTM}
\end{equation}
As long as this CTM is non empty, 
we prove that  any two 
arbitrary  events $I$
and $F$ within this block 
can be joined by a piecewise smooth, future (resp.\ past) directed timelike curve starting from the event $I$
and terminating at $F$.

We begin by noting that in this block,
the vector field $V$ obeys
$g(V,V)=-\frac {\rho^{2}\Delta(r)}{I^{2}}$
and thus 
identifies the future part of the light cone
 (points on the ring singularity are not considered as part of the spacetime). 
Moreover the axial Killing field satisfies:
\begin{equation}
g(\xi_{\varphi},\xi_{\varphi})=g_{\varphi\varphi}=\frac {\sin^{2}\vartheta}{I^{2}\rho^{2}}
[{\hat {\Delta}(\vartheta)}(r^{2}+a^{2})^{2}-{\Delta(r)}a^{2}\sin^{2}\vartheta]
\label{Eq:PHI}
\end{equation}
and upon using  (\ref{Eq:IR}) we find:
\begin{equation}
g(\xi_{\varphi},\xi_{\varphi})=\frac {\sin^{2}\vartheta}{I^{2}\rho^{2}}\left[(r^{2}+a^{2})(r^{2}+a^{2}\cos^{2}\vartheta)
+2mra^{2}\sin^{2}\vartheta+
\frac {\Lambda a^{2}}{3}(r^{2}+a^{2})[(r^{2}+a^{2})\cos^{2}\vartheta+r^{2}\sin^{2}\vartheta]\right]
\label{Eq:PHII}
\end{equation}
Since in this block, $r$ takes negative values, the term in the square bracket can be negative.
Indeed, 
evaluating the right hand side on the $\vartheta=\frac{\pi}{2}$ equatorial plane,
we find
\begin{equation}
g(\xi_{\varphi},\xi_{\varphi})=\frac {1}{I^{2}}[(r^{2}+a^{2})(1+\frac {\Lambda a^{2}}{3})+\frac {2ma^{2}}{r}]
\label{Eq:CVP}
\end{equation}
and thus for sufficiently small negative $r$,
 $g(\xi_{\varphi},\xi_{\varphi})<0$,
i.e $\xi_{\varphi}$ becomes timelike. 
By continuity arguments, the CTM defined
in (\ref{Eq:CTM})
is a non empty spacetime region.
Since the orbits of $\xi_{\varphi}$ are closed curves  around the rotation axis,
therefore near the ring singularity and for $r<0$, causality violations take place in the sense that 
at any event $q$ such that $g(\xi_{\varphi},\xi_{\varphi})_{q}<0$ there exists a closed timelike curve through $q$.\\ 

We now explore consequences of this property
and we begin by considering 
two arbitrary events $(I,F)$ within this block coordinatized according to
 $I=( t_{i},r_{i},\vartheta_{i},\varphi_{i})$,
  $F=(t_{f},r_{f},\vartheta_{f}, \varphi_{f})$.

At first we construct a  future directed timelike curve 
that begins at $I$
and terminates at an event lying on the equatorial plane\footnote{In this section, by the term equatorial plane of the CTM
we mean the collection of events coordinatized according to: $(t,r, \frac {\pi}{2}, \varphi)$ with $-\infty<t<\infty$, $\varphi$ varying in the usual range, while $r$ is negative and is chosen to satisfy the restriction:
$g(\xi_{\varphi},\xi_{\varphi})_{\vartheta=\frac {\pi}{2}}<0$.}
 of the CTM. To show that such a curve exists, 
 we consider first a 
 smooth 
non intersecting curve $\gamma(\lambda)=(r(\lambda), \vartheta(\lambda)), \lambda \in [0, 1]$,
on the 
$\left\{ (r,\vartheta) \right\}$-plane
that starts from 
$(r_{i}, \vartheta_{i})$, i.e.  for $\lambda=0$ obeys $(r(0), \vartheta(0))=(r_{i}, \vartheta_{i})$ 
while for $\lambda=1$ it
terminates at some point
$(r, \frac {\pi}{2})$
 on the equatorial plane of the CTM i.e.\
 $(r(1),\vartheta(1))=(r, \frac {\pi}{2})$.
 Such a curve always exists
 and its tangent vector  $\dot \gamma$  satisfies
\begin{equation}
g(\dot {\gamma},\dot {\gamma})={\rho(\lambda)^{2}}
[\frac {(\dot r(\lambda))^{2}}{\Delta(r(\lambda))}
+\frac {(\dot \vartheta(\lambda) )^{2}}{ {\hat \Delta}(\vartheta(\lambda))}],\quad \lambda \in [0, 1].
\label{Eq:MT}
\end{equation}
Smoothness of 
$\gamma$ combined with the 
compactness of 
the domain $[0,1]$  imply 
that the right hand side is bounded on $[0,1]$. 
Utilizing the
 integral curves of the vector field $V=(r^{2}+a^{2})\frac {\partial}{\partial t} +a\frac {\partial}{\partial \varphi}$
we now define a new curve:
\begin{equation}
\hat \gamma(\lambda)=(\gamma(\lambda),\varphi_{i}+Aa\lambda, t_{i}+At(\lambda)),\quad
 \lambda \in [0, 1],\quad A>0 
\label{Eq:ICC}
\end{equation}
with $A$ a constant and  $t(\lambda)$ satisfying $\dot t(\lambda)=r^{2}(\lambda)+a^{2}$. 
This new curve is smooth and 
its
tangent vector $\dot {\hat \gamma}$ satisfies
 \begin{equation}
 \dot {\hat \gamma}=\dot \gamma+AV,\quad g(\dot {\hat \gamma},\dot {\hat \gamma})=
  g(\dot {\gamma},\dot {\gamma})-\frac {A^{2}\Delta(r)\rho^{2}}{I^{2}},\quad g(\dot {\hat{\gamma}}, V)=Ag(V,V)<0
 \label{Eq:IM}
\end{equation}
and thus by choosing $A$ sufficiently large,  $\hat {\gamma}$ is  timelike and future pointing.
Moreover, it  begins  
at  $I= ( t_{i},r_{i},\vartheta_{i},\varphi_{i})$
and terminates at  the event $( t_{i}+At(1), r, \frac {\pi}{2}, \varphi_{i}+Aa)$
which  lies on the equatorial plane of the CTM.\\
 
 By interchanging 
$I=( t_{i},r_{i},\vartheta_{i},\varphi_{i})$ for
$F=(t_{f},r_{f},\vartheta_{f},\varphi_{f})$
and motivated by the structure of the curve
$\hat {\gamma}$ in 
(\ref{Eq:ICC}), 
we consider the curve
 \begin{equation}
\hat \gamma_{1}(\lambda)=(\gamma_{1}(\lambda),\varphi_{f}-Aa\lambda, t_{f}-At(\lambda)),\quad
 \lambda \in [0, 1],\quad A>0
\label{Eq:I2C}
\end{equation}
where here $\gamma_{1}(\lambda)=(r_{1}(\lambda), \vartheta_{1}(\lambda))$ 
satisfies:
$(r_{1}(0), \vartheta_{1}(0))=(r_{f}, \vartheta_{f})$ and
$(r_{1}(1),\vartheta_{1}(1))=(r_{1}, \frac {\pi}{2})$
subject to the restriction that $(r_{1}, \frac {\pi}{2})$ lies on the equatorial plane of the CTM.
This 
$\hat \gamma_{1}$ is timelike but it is past directed 
and joins 
$F=(t_{f},r_{f},\vartheta_{f},\varphi_{f})$ to
the event $( t_{f}-At(1), r_{1}, \frac {\pi}{2}, \varphi_{f}-Aa)$  lying on the equatorial plane of 
the CTM. For later use note that by reversing the parametrization in (\ref{Eq:I2C})
the resulting curve
is a future pointing timelike curve which 
 joins 
$( t_{f}-At(1), r_{1}, \frac {\pi}{2}, \varphi_{f}-Aa)$ to 
the event $F=(t_{f},r_{f},\vartheta_{f},\varphi_{f})$.\\

We now prove the following  property 
of the CTM: any two arbitrary events 
  $A$  and $B$
on the equatorial  plane of the CTM
can be joined by a future directed timelike curve.
We prove this property in two steps. Firstly we consider 
the special events 
 $A=( t_{0},r_{0},\frac{\pi}{2},\varphi_{0})$  and $B=( t_{0},	\hat r_{f},\frac{\pi}{2},\hat \varphi_{f})$
on the equatorial plane of the CTM.
Since 
$g(\xi_{\varphi},\xi_{\varphi})<0$
within the CTM, we show that these special events $A$ and $B$ can be joined by a 
 timelike future directed curve.\\ 
 To show this, we consider
the curve:
 \begin{equation}
\hat \gamma_{2}(\lambda)=(t_{0}, r(\lambda), \frac {\pi}{2}, 
\varphi_{0}+(\hat \phi_{f}-\phi_{0}+2\pi n)\lambda),\quad
 \lambda \in [0, 1] 
\label{Eq:NIC}
\end{equation}
where the smooth function  $r(\lambda)$ satisfies $r(0)=r_{0}$,
 $r(1)=\hat r_{f}$
and $n$ is for the moment an arbitrary positive integer.
For this curve, its  
 tangent vector $\dot {\hat \gamma}_{2}$ satisfies
 \begin{equation}
g( \dot {\hat \gamma}_{2}, \dot {\hat \gamma}_{2})=
 {\rho(\lambda)^{2}}
\frac {(\dot r(\lambda))^{2}}{\Delta(r(\lambda))}+
 (\hat \varphi_{f}-\phi_{0}+2\pi n)^{2}
 g(\xi_{\varphi},\xi_{\varphi}) ,\quad g(\dot {\hat{\gamma_{2}}}, V)=
  (\hat \varphi_{f}-\varphi_{0}+2\pi n) g(\varphi, \varphi)<0.
\label{Eq:MAU}
\end{equation} 
 and since $g(\xi_{\varphi},\xi_{\varphi})<0$, by choosing $n$ sufficiently large,
it follows that the resulting $\hat \gamma_{2}$ is timelike and future directed
 joining
 $A=( t_{0},r_{0},\frac{\pi}{2},\varphi_{0})$ to the event  $B=( t_{0},\hat r_{f},\frac{\pi}{2},\hat \varphi_{f})$.\\

 We now prove the second step and for this part we consider again two arbitrary 
 events 
 $A=( t_{0},r_{0},\frac{\pi}{2},\varphi_{0})$,
 and $B=( \hat t_{f},\hat r_{f},\frac{\pi}{2},\hat \varphi_{f})$  
  where now 
$T=\hat t_{f}-t_{0}$ is arbitrary. We show again that these events can  be joined
by  a 
 timelike, future directed curve. To show this, we appeal to the previous step and consider
 first the
curve $ \hat \gamma_{2}(\lambda)$ in (\ref{Eq:NIC})
which joins 
 $A=( t_{0},r_{0},\frac{\pi}{2},\varphi_{0})$
to  the intermediate event $C=(t_{0},\hat r_{f},\frac{\pi}{2},\hat \varphi_{f})$.
Furthermore, we introduce two new 
curves 
$\delta_{\epsilon}$ via
 \begin{equation}
\delta_{\epsilon}(\lambda)= (t_{0}+\epsilon\lambda,\hat r_{f}, \frac{\pi}{2}, \hat \varphi_{f}-b\lambda ),\quad
\lambda \in [0, T],\quad \epsilon=\pm 1
\label{Eq:NNC}
\end{equation} 
which 
join $C=(t_{0},\hat r_{f},\frac{\pi}{2},\hat \varphi_{f})$
to $B=( t_{0}+\epsilon T,\hat r_{f},\frac{\pi}{2},\hat \varphi_{f})$
provided we take $b=\frac {2\pi n}{T}$ where $n$ is a non zero integer.
For these curves, the
tangent vector $\dot { \delta_{\epsilon}}=\epsilon \frac{\partial }{\partial t}-b\frac{\partial }{\partial \varphi}$ satisfies:
\begin{equation}
g( \dot { \delta_{\epsilon}}, \dot {\delta_{\epsilon}})=\epsilon^{2}g(\xi_{t},\xi_{t})
   -2\epsilon bg(\xi_{t},\xi_{\varphi})
      +b^{2}g(\xi_{\varphi},\xi_{\varphi})
\label{Eq:TDI}
\end{equation}  
 \begin{equation}
g( \dot { \delta_{\epsilon}}, V)=
\epsilon (r^{2}+a^{2})g(\xi_{t},\xi_{t})+\epsilon ag(\xi_{t},\xi_{\varphi})
-b[(r^{2}+a^{2})g(\xi_{t},\xi_{\varphi})
+ag(\xi_{\varphi},\xi_{\varphi})].
\label{Eq:FDI}
\end{equation}  
From 
(\ref{Eq:TDI}), it is seen that by
 taking  $b^{2}$ large enough,
both of the curves $\delta_{\epsilon}$ are timelike.
Moreover working out the right hand side of
(\ref{Eq:FDI}) by evaluating the covariant components
of $g$ on the equatorial plane using  (\ref{Cov1},\ref{Cov2}), we find
\begin{equation}
g( \dot { \delta_{\epsilon}}, V)=-\frac {\Delta(r)}{I^{2}}(\epsilon+ab)
\label{Eq:FFI}
\end{equation}  
and since $\Delta (r)>0$, therefore
the curve $\delta_{1}$ which joins
$A=( t_{0},r_{0},\frac{\pi}{2},\varphi_{0})$,
to  $B=( \hat t_{f},\hat r_{f},\frac{\pi}{2},\hat \varphi_{f})$ 
 with $t_{f}=T+t_{0}$, is timelike and future directed.
 On the other hand, the curve  
 $\delta_{-1}$
that joins\footnote{It is worth pointing out here an important difference between the curves $\delta_{\pm1}$
introduced above. While both are timelike and future pointing
note that 
$\delta_{1}(t)>0$ implying that $t$ increases along 
$\delta_{1}$ while for the case of $\delta_{-1}$ we have
 $\delta_{-1}(t)<0$, i.e the coordinate $t$ decreases as one moves along $\delta_{-1}$.
 It is this property of the curve
 $\delta_{-1}$ which is responsible for traveling $backward$ $in$
 $time$. An observer following  $\delta_{-1}$,
 while moving towards  the future, finds as a consequence of 
  $\delta_{-1}(t)<0$ that the value of the Boyer-Lindquist $t$
 steadily reduces.} 
$A=( t_{0},r_{0},\frac{\pi}{2},\varphi_{0})$,
to $B=( \hat t_{f},\hat r_{f},\frac{\pi}{2},\hat \varphi_{f})$ 
 with $t_{f}=t_{0}-T$, is timelike and future directed
 provided we choose $b>a^{-1}$. In any case, the events
 $A=( t_{0},r_{0},\frac{\pi}{2},\varphi_{0})$
 and 
 $B=(\hat  t_{f},\hat r_{f},\frac{\pi}{2},\hat \varphi_{f})$ 
 can always be joined by a future directed timelike curve lying within the equatorial plane of 
 the CTM  irrespective of whether
 $T=\hat t_{f}-t_{0}$ is positive, negative or zero. 

Clearly, this conclusion holds for the choices:
 $A=(t_{i}+At(1),r, \frac {\pi}{2}, \varphi_{i}+Aa)$
and 
$B=(t_{f}-At(1), r_{1}, \frac {\pi}{2}, \varphi_{f}-Aa)$.
Accordingly, these  two events can be 
 joined by
a timelike future directed curve lying on the CTM
and this conclusion almost proves the proposition.
Indeed starting 
from the event $I=( t_{0},r_{0},\vartheta_{0},\varphi_{0})$,
the future directed timelike curve
in (\ref{Eq:ICC}) joins $I$ to the event 
$A= (t_{i}+At(1), r, \frac {\pi}{2}, \varphi_{i}+Aa)$ on the equatorial plane of the CTM,
while 
the timelike and future directed curve $\hat \gamma_{2}$  in (\ref{Eq:NIC}) combined with one of the 
 timelike and future directed curves
$\delta_{1}$ or $\delta_{-1}$ connects 
$A= (t_{i}+At(1), r, \frac {\pi}{2}, \varphi=\varphi_{i}+Aa)$ to
$B=(t_{f}-At(1), r_{1}, \frac {\pi}{2}, \varphi_{f}-Aa)$.
Finally,  the future directed timelike  $\hat \gamma_{1}$
in (\ref{Eq:I2C}) (with 
reversed parametrization)   
connects this $B$ to the event $F=(t_{f},r_{f},\vartheta_{f}, \varphi_{f})$.
Thus the non empty property of the CTM enables us to
connect 
the arbitrary events 
$I$ and $F$
by a (piecewise smooth) timelike, future directed curve that starts  from $I$ and terminates at $F$.\\

To complete the proof of the proposition, we need to show that 
the events $I$ and $F$ can also be  
connected by a timelike curve which is past directed. The proof
of this claim can proceed along 
the same lines as for the case of the future curve
that joins $I$ to $F$,
but here we follow a shortcut that avoids this procedure.
The existence of a timelike past directed 
curve starting from $I$ and terminating at $F$ can be inferred 
by interchanging the roles of $I$ and $F$
in the previous proof.
Accordingly, there exists
 a future directed 
timelike curve which originates at $F$ and terminates at $I$. Hence by a parametrization reversal 
this curve becomes a past directed timelike curve
from $I$ to $F$
and this conclusion completes the proof of the proposition.\\
 
 In the limit that $\Lambda\to0$,
we recover Carter's
results for the case of Kerr.
 The Boyer-Lindquist block  that contains
the ring singularity is a vicious set.
Carter arrived at this conclusion by appealing to the properties 
of the two dimensional transitive Abelian isometry 
 group acting on the background Kerr (or Kerr-Newman) spacetime.
 Even though his method can probably be adapted to
 cover the case of a Kerr-de Sitter,
 in this work we have chosen an alternative proof which, though pedestrian,
 nevertheless makes clear 
 the role played by the CTM in destroying any notion of causality.
 Our approach is along the lines of a proof outlined in 
 ref.\cite{Neil} although in the present work the background is
 different
from the one in ref.\cite{Neil}, 
 and we use a different representation of
the (highly non unique) family of causal curves that join the events
under consideration. Also, Chrusciel in \cite{Chr2} discusses qualitatively
properties of the CTM for the case of Kerr background.\\

The proof of the proposition
(\ref{Prop:MfB}) shows that even a tiny non empty CTM converts the entire block
to a vicious set where any notion of causality is lost. Through  any event
on this block, the CTM generates a closed timelike curve through this event
(for some properties of the vicious regions of a Kerr spacetime see
for instance (\cite{Y1}, \cite{Y2}).
\\
 
\section{Discussion}
In this work, the causality properties of the family of Kerr-de Sitter 
spacetimes have been worked out
and the main conclusions are summarized 
in the three propositions proved in the previous section.
 Even though our emphasis has focused on the family
of the Kerr-de Sitter spacetimes describing a black hole enclosed within a pair 
 of cosmological horizons (for a discussion supporting this interpretation see
  \cite{GibHaw1},\cite{Mat1}),
 the  propositions of the last section remain valid whenever
  the equation $\Delta(r)=0$ admits double roots of roots of higher multiplicity.
  For instance, for the case
  where
 $\Delta(r)=0$ admits only two real roots $r_{1}<0<r_{2}$,  the underlying spacetime
  describes a
  singular ring enclosed within a pair of cosmological horizons.
  The region between the cosmological horizons is a vicious set, while
  the asymptotic de Sitter like regions are causally  well
 behaved. This behavior is to be contrasted 
 to the case of a Kerr spacetime describing a naked singularity where
 the asymptotic region fails to be causally well behaved.\\

In summary, this work shows that the causality violating regions in a Kerr-de Sitter  spacetime
 consist of the disjoint union of the Carter's blocks
 that contain the ring singularity.
 It should be stressed however that this 
 conclusion assumes  that the global structure of the underlying 
 spacetime is the one shown in Fig.2.
 If however, one of the  $r\to -\infty$ asymptotic regions 
 is to be identified with an $r\to \infty$ region (see Fig.2 and comments in the caption of this figure)
 then 
 the change in the connectivity properties of the 
 underlying manifold leads to the appearance of closed causal curves through the 
 asymptotic regions. These causality violations are of the same nature as those
 encountered 
 whenever 
 different asymptotic regions of a Kerr \cite{Car4} or a  Reissner-Nordstrom
spacetime  \cite{Car5} are identified.
As  pointed out by Carter \cite{Car3},
the  causality violating curves
are not homotopic reducible  to a point 
and thus they can be eliminated by moving
to a suitable covering spacetime manifold (see discussion in \cite{Car3}).
However,
the causality violations occurring in a Kerr-de Sitter spacetime within the 
blocks that contain the ring singularity
is of a different nature since 
the causality violating curves cannot  
be removed by moving to a suitable covering space.
This type of causality violation
also occurs for the Kerr or Kerr-Newman family 
and furthermore there exist other solutions of 
 Einstein's equations that 
exhibit the same behavior.
The best known example is provided by G\"odel's\footnote{Although
G\"odel's solution \cite{God} seems to be the best known example of a spacetime violating causality,
chronologically it is not the first constructed solution of Einstein's equations that exhibits this property.
In 1937, van Stockum \cite{vanS} published a solution of Einstein's equations with source a 
 rapidly rotating, infinitely long, dust cylinder
and showed that this spacetime admits closed timelike curves.
A re-examination of the causality properties of the van 
Stockum solution
has been presented in  \cite{Tip1}.} solution \cite{God}. The solution admits closed timelike curves that are homotopic reducible  to
a point and thus in the G\"odel universe a non trivial causality violation takes place
(for an introduction to G\"odel's solution see for instance \cite{HE}).
For a review of spacetimes exhibiting non trivial causality violations
see \cite{LOB}.
The ref. 
\cite {TTTM} discusses properties 
of closed causal curves, time travel and time machines.\\

For the moment, there is no consensus 
regarding the role of spacetimes
exhibiting
non trivial causality violations
in describing reality.
 For instance
Hawking  in \cite{Haw1} 
presents evidence that quantum effects probably eliminate
the appearance of closed causal curves
and he introduced  the Chronology Protection Conjecture
stating: The  laws of physics do not allow
the appearance of closed timelike curves.
However other authors,
notably Thorne and collaborators\footnote{In ref.\cite{Tho1},
it is asked whether the laws of physics permit the creation of wormholes in a universe whose spatial sections initially are simply connected. If the laws indeed allow the formation of wormholes, then
the appearance of closed timelike curves (and also violation of 
the weak energy condition) is unavoidable. For a proof of the former property
see \cite{Haw1}, while for the latter see  \cite{Tip2}. } 
 take a different attitude towards causality violating spacetimes.
Rather than considering them as an anomaly, 
they take the viewpoint 
that it is prudent to investigate thoroughly their consequences.
For instance, in  \cite{Ori1} it is argued that closed timelike curves 
may be generated by 
matter satisfying the weak energy condition,
a situation to be contrasted with the
spirit of the 
Chronology Protection Conjecture.
Irrespective, however, of the attitude 
that one takes towards spacetimes violating causality,
clearly it is helpful to have a good supply of exact solutions
 of Einstein's equations
exhibiting causality violating regions.
This work 
added to this compartment another family 
of such solutions, 
namely the family of Kerr-de Sitter spacetimes.
 \\

We finish this paper by mentioning that ever since cosmological observations
suggest that we live in Universe with accelerating expansion, studies of solutions of Einstein's
equations with a non vanishing cosmological constant $\Lambda$ are becoming the
focus of intense investigations. Some current results dealing with classical and quantum aspects of 
Kerr-(anti)-de Sitter can be found in refs.\cite{X1},\cite{X2},\cite{X3},\cite{X4}.

\section{Appendix}
In this Appendix, we remind the reader of a few basic notions of causality theory
(a more elaborate discussion can be found in 
 \cite{HE}, \cite{Wald}). We recall that for any physically relevant spacetime $(M, g)$,
besides the standard requirements that $M$ ought to be smooth, connected, Hausdorff and paracompact,
it is further required that $(M, g)$ to be time orientable
and causally well behaved.
Causally well behaved means that $(M,g)$
is minimally 
 causal (resp.\ chronological) 
according to the definition:
\begin{definition} 
  A time orientable spacetime $(M,g)$ is said to be causal (resp.\ chronological), if 
 admits  no causal (resp.\ timelike) closed curves.
 \end{definition}
The absence of closed timelike curves from any physically relevant $(M,g)$
 is required to be a stable property of $(M,g)$ in the sense that any small 
 perturbations of  the background metric $g$
 should not lead to the appearance of closed 
  causal curves. This additional  requirement leads
to the notion of
stable causality according to the definition: 
 
  \begin{definition} 
  A time orientable spacetime $(M,g)$ is stably causal if  there exists a continuous timelike vector field $t$
 such that the spacetime $(M,\hat g)$ with  $\hat g=g-\hat t \otimes \hat t$  possesses no closed timelike curves, (here 
 the covector $\hat t$ is  defined by:
$\hat t=g(t, ) )$.
\end{definition}
A very useful criterion
 guaranteeing that a given $(M,g)$ is stably causal 
is expressed by 
the following theorem  \cite{HE},\cite{Wald}:
\begin{theorem}
A time orientable $(M,g)$ is stably causal if and only if there
exists a differentiable  function $\tau$ $($often referred as the 
time function$)$ such that 
$\nabla \tau$  is a past directed timelike vector field.
\end{theorem}
Clearly, if $(M,g)$ admits  a function $\tau: M \to R$ with these properties,
then no closed timelike curves can occur, since 
for any future directed timelike curve $\gamma$ with tangent vector field $X$,  the inequality
$0<g(X, \nabla \tau)=X(\tau)$ 
implies that $\tau$ is strictly increasing along 
this $\gamma$.
Therefore, under the hypothesis of the theorem, there exist no closed timelike curves in this  $(M,g)$.
The proof of the converse is more involved but it can be found
 in \cite{HE},\cite{Wald}.\\


\acknowledgments

It is a pleasure to thank the members of the relativity group at the Universidad Michoacana for stimulating discussions.
Special thanks are due to O. Sarbach for his constructive criticism,
J. Felix Salazar for discussions and for his help in drawing the diagrams
and F. Astorga for comments on the manuscript.
The research was supported in part by CONACYT Network Project 280908 Agujeros Negros y Ondas 
Gravitatorias and by a CIC Grant from the Universidad Michoacana.\\

\end{document}